\documentclass[aps,preprint,epsfig,rotate]{revtex4}
\usepackage{graphicx}
\usepackage{bm}
\usepackage{epsfig}

\input epsf
\begin{document}

\title{Hyperfine structure splitting for the triplet $2^{3}S(L 
       = 0)-$state in the two-muonic helium ${}^{3}$He 
       $\mu_{2}$ atom}       

\author{Alexei M. Frolov}
\email[E--mail address: ]{alex1975frol@gmail.com}

\affiliation{Department of Applied Mathematics \\
 University of Western Ontario, London, Ontario N6H 5B7, Canada}

\date{November 25, 2026}

\begin{abstract}

The hyperfine structure splitting for the triplet $2^{3}S(L = 0)-$state 
in  the two-muonic helium ${}^{3}$He $\mu_{2}$ atom is determined to 
high accuracy. Such a structure arises as a result of hyperfine (or 
spin-spin) interaction between the triplet two-muonic shell and non-zero 
spin of the central helicon nucleus. Our result for the hyperfine 
structure splitting is $\Delta \approx 2.5811701551505 \cdot 10^{8}$ 
MHz, where the standard uncertainty has been evaluated as $\pm 100$ 
Hz.  
 
\end{abstract}

\maketitle
\newpage

The hyperfine structure splitting for the triplet $2^{3}S(L = 0)-$state 
in the two-electron helium ${}^{3}$He atom (or helicon atom, for short) 
was predicted a long time ago (see, e.g., \cite{Hughes1}, \cite{Hughes2}). 
This problem has been investigated in \cite{BS}, which also contains a 
number of useful references on this subject published before 1956. 
Accurate and highly accurate measurements of this hyperfine structure 
splitting started in the middle of 1960's (see, e.g., \cite{RosPip} and 
references therein). Currently, the overall accuracy of such measurements 
is very high and constantly increases. Our evaluation of the hyperfine 
structure splitting for the triplet $2^{3}S(L = 0)-$state in the helicon 
atom (or ${}^{3}$He atom) can be found, e.g., in \cite{Fro2007} and 
\cite{Fro2016}. In this study we want to evaluate the hyperfine structure 
splitting for the bound triplet $2^{3}S(L = 0)-$state in the two-muonic 
${}^{3}$He $\mu_2$ atom. This problem is of great interest in modern 
atomic physics due to its obvious similarity to the analogous problem for 
the ordinary, two-electron helicon atom ${}^{3}$He. 

In general, the same statement can be repeated for other bound state 
properties of various, two-muonic light atoms, including atoms of helium,  
lithium, etc. Currently, for a number of light two-muonic atoms and ion 
there are a few dozens of interesting, but unsolved problems. For 
instance, it is possible to investigate variation of atomic properties 
in the series from the ordinary helicon atom to the helicon muonic atom 
and then to the helicon two-muonic atom. It is clear that such an 
investigation can  produce a large volume of useful information. Note 
also that soon it will be possible (as I hope) to create the two-muonic 
atoms experimentally. In reality, the two-muonic helicon atoms can be 
created and observed in the same experiments which were used earlier 
for creation of the helium-muonic ${}^{3}$He $\mu \; e$ and ${}^{4}$He 
$\mu \; e$ atoms (see, e.g., \cite{Hughes} and \cite{PRL}). However, the 
overall outcome of such experiments for the two-muonic helium-3 atoms 
will be significantly lower. Preliminary and quite optimistic evaluations 
indicate that for each 130 - 170 helium-muonic atoms created in reality, 
we can also observe one two-muonic helium atom. In reality, such a ratio 
can be 1:1000 and even lower. Furthermore, the main question in these 
experiments can be formulated in the form: how can we prove that this 
very compact two-muonic system is a true two-muonic helium atom (helicon) 
which is, e.g., in its triplet $2^{3}S-$state? In general, it can be done 
by using our theoretically predicted hyperfine structure splitting for 
the triplet $2^{3}S(L = 0)-$state of the two-muonic helicon ${}^{3}$He 
$\mu_2$ atom. 
 
Let us derive the explicit formula for the hyperfine structure splitting 
in the bound triplet $2^{3}S(L = 0)-$state of the two-muonic helicon 
atom ${}^{3}$He $\mu_2$. The well known formula for the Hamiltonian 
$\hat{H}_{HSS}$, which describes similar hyperfine structure splitting 
in the bound triplet $2^{3}S(L = 0)-$state of the usual two-electron 
helicon atom ${}^{3}$He (see, e.g., \cite{Jacks} - \cite{LLQM} and also 
\cite{Fro2016}), is written in the form:
\begin{eqnarray}
 \hat{H}_{HSS} = \frac{8 \pi \alpha^{2}}{3} \; g_{N} \; g_{e} \; 
 \Bigl(\frac{m_e}{M_p}\Bigr) \; (c \mu_{B})^{2} \; \langle \delta( 
 {\bf r}_{N e} ) \rangle \; \Bigl( {\bf I}_{N} \cdot {\bf S}_{e} 
 \Bigr) \; \; \; , \; \; \; \label{HSS}
\end{eqnarray}
where $\alpha = 7.2973525693 \cdot 10^{-3}$ is the fine structure constant, 
$c$ is the speed of light in vacuum, $\mu_B = \frac{\hbar e}{2 c m_e}$ is 
the Bohr magneton, while $\mu_N = \frac{\hbar e}{2 c M_P} = \mu_B \Bigl( 
\frac{m_e}{M_p} \Bigr)$ is the nuclear magneton \cite{Bmagnet}, where $M_p$ 
= 1836.15267343 $m_e$ is the proton mass. In Eq.(\ref{HSS}) and everywhere 
below in this study the notation $m_e$ designates the electron mass-at-rest. 
The actual mass of the helion atomic nucleus is 5495.88528007 $m_e$, while 
the muon mass-at-rest is 206.7682830 $m_e$. In this study all numerical 
values of physical constants and particle properties are taken from 
\cite{NIST}. Also, in Eq.(\ref{HSS}) the notation $\langle \delta( {\bf 
r}_{N e} ) \rangle$ stands for the expectation value of the electron-nucleus 
delta-function which must be taken in this equation in atomic units, where 
$\hbar = 1, m_e = 1, \mid e \mid = 1$ and $4 \pi \epsilon_0 = 1$. In atomic 
units the speed of light $c$ equals $\alpha^{-1} \approx 137.035999081...$, 
i.e., it is exactly coincides with the inverse fine structure constant. 

Also, in Eq.(\ref{HSS}) the factor $c \mu_{B} = \frac{\hbar \; e}{2 m_e}$ 
is the product of speed of light in vacuum and Bohr magneton. In atomic 
units this product equals $\frac12$ exactly, while the Bohr magneton is 
$\mu_{B} = \frac{\hbar \; e}{2 m_e c} = \frac{\alpha}{2}$. Note that in 
atomic units the Hamiltonian $\hat{H}_{HSS}$ takes the form
\begin{eqnarray}
 \hat{H}_{HSS} = \frac{\pi \alpha^{2}}{3} \; g_{N} \; g_{e} \; 
 \Bigl(\frac{m_e}{M_p}\Bigr) \; \langle \delta( {\bf r}_{N e} ) 
 \rangle \; \Bigl[ F (F + 1) - I_N (I_N + 1) - S_{e} (S_{e} + 
 1) \Bigr] \; \; , \; \; \; \label{HSSAU}
\end{eqnarray}
where ${\bf F} = {\bf I}_{N} + {\bf S}_{e}$ is the total electron-nucleus 
spin vector, while the symbols ${\bf I}_{N}$ and ${\bf S}_{e}$ 
Eq.(\ref{HSSAU}) designate the spin-vector of the helium-3 nucleus and 
combined spin-vector of the two atomic electrons, respectively, i.e., 
${\bf S}_{e} = {\bf s}_{1} + {\bf s}_{2}$. In general, the absolute value 
of the nuclear spin $I_{N} = \mid {\bf I}_{N} \mid$ is always integer 
and/or semi-integer. For the bound triplet $2^{3}S(L = 0)-$state in the 
helicon atom one finds $S_e = 1, I_N = \frac12$. This leads to the 
following values of $F = \mid {\bf F} \mid$: $F_{q} = \frac32$ (quartet 
state) and $F_{d} = \frac12$ (doublet state). In Eq.(\ref{HSS}) we apply 
the electron and helicon spin $g-$factors which are the dimensionless 
ratios of the electron's and nucleus's magnetic moments to their 
intrinsic spins, respectively. In particular, for the electron we have 
$g_{e}$ = - 2.00231930436256, while for the helicon nucleus one finds 
$g_{N} = \frac{{\cal M}_{N}}{I_N}$ = - 4.255250615. In other words, $g_N 
= \frac{{\cal M}_N}{I_N}$, where ${\cal M}_N$ = - 2.127625307 (in nuclear 
magnetons $\mu_N = \mu_{B} \Bigl(\frac{m_e}{M_P}\Bigr) \Bigr)$ to the 
nuclear spin ${I}_N$ which equals $\frac12$ for the helicon nucleus. 

The eigenvalues of the $\hat{H}_{HSS}$ matrix from Eq.(\ref{HSSAU}), 
coincide with the corresponding hyperfine structure splittings. 
However, all these eigenvalues will be expressed in atomic units, 
while traditionally (since the early 1950's \cite{Rams}, \cite{Heine}) 
the hyperfine structure splittings in atoms and molecules are always 
expressed in Hertz (or Hz), or in Mega Hertz (or MHz, for short), where 
1 MHz = $1 \cdot 10^{6} \;$ Hz. To re-calculate the energies expressed 
in atomic units into MHz we have to apply the conversion factor $Ry_{e}$ 
= 6.5796839205016$\cdot 10^{9}$ MHz $\cdot \; (a.u.)^{-1}$, where the 
factor $Ry_{e}$ numerically coincides with the atomic (or electron) 
Rydberg constant. Now, our formula, Eq.(\ref{HSSAU}), is reduced to its 
final form: 
\begin{eqnarray}
 \hat{H}_{HSS} = \frac{\pi \alpha^{2}}{3} \; g_{N} \; g_{e} \; 
 \Bigl(\frac{m_e}{M_p}\Bigr) \; \langle \delta( {\bf r}_{N e} ) 
 \rangle \; \Bigl[ F (F + 1) - I_N (I_N + 1) - S_{e} (S_{e} + 
 1) \Bigr] \; Ry_{e} \; . \; \; \; \label{HSSAUE}
\end{eqnarray}
This formula allows one to reproduce our earlier result obtained in 
\cite{Fro2007}, which equals 6740.4521391 MHz (or 6740.4529188 MHz 
with the recent physical constants). The corresponding experimental 
result is 6739.701177 MHz \cite{RosPip}. 

Now, let us transform the `atomic' formulas, Eqs.(\ref{HSS}) - 
(\ref{HSSAUE}) to the `muonic' forms which are appropriate for 
applications to the two-muonic atom(s). First, we multiply the right 
hand side of Eq.(\ref{HSSAU}) by the factor $f$, which equals unity 
and is represented in the form 
\begin{eqnarray}
 f = \Bigl( \frac{m_{\mu}}{m_e} \Bigr)^{2} \Bigl( 
 \frac{m_e}{m_{\mu}} \Bigr)^{2} = 1 \; \; \; . \; 
 \; \; \label{Factemu}
\end{eqnarray} 
After a few simple transformations we reduce this formula to the 
new form 
\begin{eqnarray}
 H_{HSS} = \frac{8 \pi \alpha^{2}}{3} \; g_{N} \; g_{e} \; 
 \Bigl(\frac{m_{\mu}}{M_p}\Bigr) \; \Bigl[ c \mu_{B} \Bigl( 
 \frac{m_e}{m_{\mu}} \Bigr) \Bigr]^{2} \; \langle \delta( 
 {\bf r}_{N e} ) \rangle \; \Bigl(\frac{m_{\mu}}{m_e}\Bigr) \; 
 \Bigl( {\bf I}_{N} \cdot {\bf S}_{e} \Bigr) \; \; \; . 
 \; \; \; \label{HSSEMu}
\end{eqnarray} 
To simplify this formula we introduce the muonic magneton 
$\mu^{(\mu)}_{B}$, which is defined as follows 
\begin{eqnarray}
 \mu^{(\mu)}_{B} = \mu_{B} \Bigl( \frac{m_e}{m_{\mu}} \Bigr) 
 =  \frac{\hbar e}{2 c m_e} \; \Bigl(\frac{m_e}{m_{\mu}}\Bigr) 
 = \frac{\hbar e}{2 c m_{\mu}} \; \; \; . \; 
 \; \; \label{BMamu} 
\end{eqnarray} 
The explicit use of muonic magneton allows us to write the formula 
for the hyperfine structure splitting in the two-muonic helium atoms 
in the form 
\begin{eqnarray}
 H_{HSS} = \frac{8 \pi \alpha^{2}}{3} \; g_{N} \; g_{\mu} \; 
 \Bigl(\frac{m_{\mu}}{M_p}\Bigr) \; \; \Bigl[ c \; 
 \mu^{(\mu)}_{B} \Bigr]^{2} \; \langle \delta( {\bf r}_{N \mu} ) 
 \rangle \; \Bigl(\frac{m_{\mu}}{m_e}\Bigr) \; \Bigl( {\bf I}_{N} 
 \cdot {\bf S}_{\mu} \Bigr) \; \; \; . \; \; \; \label{HSSMu}
\end{eqnarray} 
where we made the three obvious replacements in Eq.(\ref{HSSEMu}): 
\begin{equation}
 g_{e} \rightarrow g_{\mu} \; \; \; , \; \; \; \langle \delta( 
 {\bf r}_{N e} ) \rangle \; \rightarrow \langle \delta( {\bf 
 r}_{N \mu} ) \rangle \; \; \; \; , \; \; \; {\bf S}_{e} 
 \rightarrow {\bf S}_{\mu}  \; \; . \; \; \; \label{replac}
\end{equation} 
where ${\bf S}_{\mu} = {\bf s}_{1} + {\bf s}_{2}$ is the combined 
spin-vector of two atomic muons, while the second replacement here 
means that the muonic-nucleus delta-function must be taken in muon 
atomic units, where $\hbar = 1, m_{\mu} = 1, \mid e \mid = 1$ and 
$4 \pi \epsilon_0 = 1$. In fact, all replacements in 
Eq.(\ref{replac}) are pretty obvious and they do not require to 
apply any additional factors. It is also clear that by making such 
replacements we simply remove all electron's properties from 
Eq.(\ref{HSSMu}) and now the arising formula can be used for 
two-muonic atoms, including the two-muonic helium-3 atom. The 
numerical value of the muonic spin $g-$factor is $g_{\mu}$ = - 
2.0023318418, since it is known from recent experiments. 

Finally, we are ready to write our formula for the hyperfine 
structure splittings in the triplet $2^{3}S(L = 0)-$state of the 
two-muonic helium ${}^{3}$He $\mu_{2}$ atom. By taking into 
account that in muon atomic units ($m.a.u.$) the muonic magneton 
equals $\frac12$, we obtain directly from Eq.(\ref{HSSMu}) the 
following (final) formula  
\begin{eqnarray}
 \hat{H}_{HSS} &=& \frac{\pi \alpha^{2}}{3} \; g_{N} \; g_{\mu} 
 \; \Bigl(\frac{m_{\mu}}{M_p}\Bigr) \; \langle \delta( {\bf 
 r}_{N \mu} ) \rangle \; \Bigl[ F (F + 1) - I_N (I_N + 1) - 
 S_{\mu} (S_{\mu} + 1) \Bigr] \; Ry_{\mu} \; , \; \label{HSSMuA}
\end{eqnarray} 
where the expectation value of the muonic-nucleus delta-function 
$\langle \delta( {\bf r}_{N \mu} ) \rangle$ value must be determined 
in the muon atomic units. Also, the conversion factor $Ry_{\mu} = 
\frac{m_{\mu} e^{4}}{\hbar^{2}} = \Bigl(\frac{m_{\mu}}{m_e}\Bigr) 
\; Ry_{e}$ = 13.60469946924824$\cdot 10^{11}$ MHz in this formula 
is the conversion factor, which numerically coincides with the 
muonic Rydberg constant. In Eq.(\ref{HSSMuA}) this factor is 
expressed in MHz $\cdot (m.a.u.)^{-1}$. It is clear that there are 
obvious similarities between the two formulas, Eq.(\ref{HSSAUE}) 
and Eq.(\ref{HSSMuA}), and these similarities allow us to determine 
the hyperfine structure splittings for the triplet $2^{3}$S-state 
in two different physical systems. Note that in the case of 
two-muonic helium ${}^{3}$He $\mu_{2}$ atom we apply the following 
numerical values $S_{\mu} = 1, I_{N} = \frac12$. The two values of 
$\mid {\bf F} \mid$ equal $F_{q} = \frac32$ for the quartet of 
states and $F_{d} = \frac12$ for the doublet of states, 
respectively. We need to determine the difference between these 
two multiple states, since in these two cases we have $\Bigl[ 
F (F + 1) - I_N (I_N + 1) - S_{\mu} (S_{\mu} + 1) \Bigr] = 1$ for 
$F_{q} = \frac32$ and $\Bigl[ F (F + 1) - I_N (I_N + 1) - S_{\mu} 
(S_{\mu} + 1) \Bigr] = - 2$ for $F_{d} = \frac12$. Note also that 
the following sum $4 \; E_{HSS}(F = \frac32) + 2 \; E_{HSS}(F = 
\frac12)$ equals zero exactly (as expected). Briefly, we can say 
that it works exactly as for the two-electron helium atom (see 
above). 

Results of our calculations can be found in Table I. As follows 
from this Table the numerical values of hyperfine structure 
splitting for the triplet $2^{3}S(L = 0)-$state in the two-muonic 
helicon ${}^{3}$He $\mu_2$ atom are now known to very good 
numerical accuracy. The final result can be written in the form
\begin{eqnarray}
 \Delta = 258117015.51505 (\pm 1 \cdot 10^{-4} ) \; \; \; 
 {\rm MHz} \; , \; \; {\rm or} \; \; \; \Delta = 
 2.5811701551505(10) \cdot 10^{8} \; \; \; {\rm MHz} \; \; . \; 
 \; \label{Result}
\end{eqnarray} 
The overall numerical accuracy of this result for $\Delta$ is 
very high and outstanding. However, it is also clear that a 
number of other corrections, e.g., the lowest-order radiative 
and QED corrections, must be taken into account. Very likely, 
that after this procedure the final accuracy of our result will 
be lower and even much lower. In general, the numerical value of 
hyperfine structure splitting $\Delta$ for the triplet $2^{3}S(L 
= 0)-$state in the two-muonic helium ${}^{3}$He $\mu_2$ atom is 
$\approx$ 38293.719817 times larger than the corresponding 
$\Delta$ value estimated above for the triplet $2^{3}S(L = 
0)-$state of the two-electron helium ${}^{3}$He atom. Such a 
comparison indicates a relative scale of $\Delta$'s in the 
triplet $2^{3}S(L = 0)-$states of the two-muonic helicon atom 
and regular two-electron helicon atom.  

Thus, we investigated the problem of hyperfine structure splitting 
for the triplet $2^{3}S(L = 0)-$state in the two-muonic helium 
${}^{3}$He $\mu_2$ atom. First, by using our new method we derived 
a number of simple and physically transparent formula which can be 
used to determine the hyperfine structure splitting $\Delta$ for 
the triplet $2^{3}S(L = 0)-$state in the two-muonic helium 
${}^{3}$He $\mu_2$ atom. Then, by using our highly accurate 
expectation values of the muon-nucleus delta-functions we 
determined the numerical values of the actual hyperfine structure 
splitting $\Delta = 2.5811701551505(10) \cdot 10^{8}$ MHz, where 
the standard uncertainty has been estimated as $\pm 100$ Hz. 

This is an important result for a very compact physical system, 
which is the two-muonic helium-3 atom. Indeed, the spatial radius 
of the two-muonic helium ${}^{3}$He $\mu_2$ atom in the triplet 
$2^{3}S(L = 0)-$state is 2.6488807487775333 $m.a.u.$ $\approx$ 
6.77921829361395$\cdot 10^{-11}$ $cm$, which is a very small size. 
In reality, if one can detect any absorption and/or emission of 
radiation with the frequency $\omega \approx$ 2.581170155 $\cdot 
10^{8}$ MHz in future experiments conducted for creation of the 
two-muonic helicon atoms, then it is possible to say that such 
atoms have indeed been created. To conclude this study we note 
that the total non-relativistic energy of the similar (bound) 
$2^{3}S(L = 0)-$state in the two-muonic helium ${}^{4}$He $\mu_2$ 
atom is $\linebreak$ $\approx$ -2.1151117000459152034591521295 
$m.a.u.$ $(N = 3842)$, but this state has no hyperfine structure 
splitting, since in this atom the nuclear spin equals zero, i.e., 
$I_N = 0$. \\


I wish to thank Prof. G.W.F. Drake (University of Windsor, Windsor, 
Ontario) who has provided sufficient computational resources, which 
were critical to finish this project. \\


\begin{center}
   {\bf Declarations} 
\end{center} 

\noindent
{\bf Author contributions}: Alexei M. Frolov contributed to the 
conceptualization (equal); formal analysis (equal); investigation 
(equal); writing—original draft (equal); and writing—review and 
editing (equal). \\

\noindent
{\bf Data Availability Statement}: This manuscript has no associated 
data or the data will not be deposited. [Author's comment: All data 
generated or analyzed during this study are included in this published 
article and cited references]. \\

\noindent
{\bf Funding Statement}: I received no funding to write/supervise 
this manuscript and conduct all related research. \\ 

\noindent
{\bf Ethical Statement}: Currently, I have no `conflicts of interests' 
which can be associated with this manuscript. \\ 


%
\newpage
\begin{table}[tbp]
   \caption{The total energies $E$ and the expectation values of the 
            muon-nucleus delta-functions $\langle \delta_{N \mu} 
            \rangle$ in muon atomic units ($m.a.u.$) and hyperfine 
            structure splitting $\Delta$ (in MHz) in the triplet 
            $2^{3}S(L = 0)-$state in the two-muonic helium 
            ${}^{3}$He $\mu_{2}$ atom$^{(a)}$. The notation $N$ 
            stands for the total number of exponential basis 
            functions used.}    
     \begin{center}
    \scalebox{1.05}{%
     \begin{tabular}{| c | c | c | c |}
      \hline\hline
  $N$  &  $E$ & $\langle \delta_{N \mu} \rangle$ & $\Delta$ (in MHz) \\ 
     \hline
 3400  & -2.096171962554133934800401995 & 1.18198188766294 & 258117015.51501 \\

 3500  & -2.096171962554133934800402061 & 1.18198188766300 & 258117015.51502 \\
 
 3600  & -2.096171962554133934800402119 & 1.18198188766303 & 258117015.51503 \\

 3700  & -2.096171962554133934800402151 & 1.18198188766313 & 258117015.51505 \\

 3800  & -2.096171962554133934800402200 & 1.18198188766318 & 258117015.51506 \\

 3840  & -2.096171962554133934800402210 & 1.18198188766310 & 258117015.51504 \\

 3842  & -2.096171962554133934800402211 & 1.18198188766312 & 258117015.51505 \\
         \hline \hline  
  \end{tabular}}
  \end{center}
  ${}^{(a)}$ In these calculations the mass of the ${}^{3}$He$^{2+}$ 
  nucleus (or helicon nucleus) equals 5495.88528007 $m_e$, while the 
  mass-at-rest of the negatively charged muon $\mu^{-}$ is 206.7682830 
  $m_e$ (or 1 $m.a.u.$) \\ 
  \end{table}


\begin{thebibliography}{01}
\bibitem{Bmagnet} Such a definition of the Bohr magneton is natural. 
Indeed, the magnetic moment of the non-relativistic electron is the 
product of elementary electric charge $e$ and a small `velocity' 
factor, which is $\Bigl(\frac{v}{c}\Bigr) \approx \alpha \approx 
\frac{1}{137}$ (in atomic units).
\end{thebibliography}
\end{document}